\documentclass[%
 aip,
 amsmath,amssymb,
reprint,%
]{revtex4-2}

\usepackage{graphicx}
\usepackage{dcolumn}
\usepackage{bm}
\usepackage{multirow}
\usepackage[utf8]{inputenc}
\usepackage[T1]{fontenc}
\usepackage{mathptmx}

\usepackage{braket}
\usepackage{graphicx}
\usepackage{amsmath,amssymb,mathtools}
\usepackage{dcolumn}
\usepackage{bm}
\usepackage{subfigure}
\usepackage{color}

\begin{document}

\preprint{AIP/123-QED}

\title{Elucidating the Molecular Orbital Dependence of the Total Electronic Energy in Multireference Problems}

\author{Jan-Niklas Boyn}
 \affiliation{The James Franck Institute and The Department of Chemistry, The University of Chicago, Chicago, Illinois 60637 USA}
\author{David A. Mazziotti}%
 \email{damazz@uchicago.edu}
\affiliation{The James Franck Institute and The Department of Chemistry, The University of Chicago, Chicago, Illinois 60637 USA}%

\date{Submitted March 5, 2022\textcolor{black}{; Revised April 18, 2022}}

\begin{abstract}
The accurate resolution of the chemical properties of strongly correlated systems, such as biradicals, requires the use of electronic structure theories that account for both multi-reference as well as dynamic correlation effects. A variety of methods exist that aim to resolve the dynamic correlation in multi-reference problems, commonly relying on an exponentially scaling complete-active-space self-consistent-field (CASSCF) calculation to generate reference molecular orbitals (MOs). However, while CASSCF orbitals provide the optimal solution for a selected set of correlated (active) orbitals, their suitability in the quest for the resolution of the total correlation energy has not been thoroughly investigated. Recent research has shown the ability of Kohn-Shan density functional theory (KS-DFT) to provide improved orbitals for coupled cluster (CC) and  M\o ller-Plesset perturbation theory (MP) calculations. Here we extend the search for optimal and more cost effective MOs to post-configuration-interaction (post-CI) methods, surveying the ability of the MOs obtained with various DFT functionals, as well as Hartree-Fock, and CC and MP calculations to accurately capture the total electronic correlation energy. Applying the anti-Hermitian contracted Schr\"odinger equation (ACSE) to the dissociation of N$_2$, the calculation of biradical singlet-triplet gaps and the transition states of the bicylobutane isomerization, we demonstrate DFT provides a cost-effective alternative to CASSCF in providing reference orbitals for post-CI dynamic correlation calculations.
\end{abstract}

\maketitle

\section{Introduction}
The computational resolution of electronic structure relies on the accurate capture of the correlation energy, which is defined as the difference between the full-configuration-interaction (FCI) and Hartree-Fock (HF) energies. The correlation energy is generally further divided into two components: static or strong correlation arising from a state that may not be described by a single Slater determinant and is hence also termed multi-reference correlation, and the remainder which is defined as dynamic correlation\cite{Reiher_CorBalance, PopleCorrelation, Correlation_Decomposition, Definition_Correlation}. While dynamic correlation is present in all electronic systems and may be well described by many single-reference methods such as coupled cluster (CC), M\o ller-Plesset perturbation theory (MP)\cite{CCMPCorRev} or even density functional theory (DFT)\cite{DFTcor1, DFTcor2}, strong correlation only arises in systems exhibiting a degeneracy or near-degeneracy of electronic states\cite{Reiher_CorBalance}. As such, multi-reference correlation plays a particularly important role in processes such as bond dissociation, and in the determination of properties of bi- or multi-radical systems, such as spin state splittings and magnetic couplings in molecules and complexes in the areas of spintronics, photonics or catalysis\cite{Birad,Birad1, Birad2, Birad3}. \\

Multi-reference correlation is commonly resolved with complete active space configuration interaction (CASCI) or CAS self consistent field (CASSCF) calculations, which resolve the strong correlation in a chosen active space\cite{CASSCF1, CASSCF2, CASSCF3, RUEDENBERG198265}. \textcolor{black}{While CASSCF calculations have proven valuable in the description of systems dominated by multi-reference correlation\cite{CASSCF1, CASSCF2},} it has been demonstrated that even in \textcolor{black}{such} systems, experimentally relevant properties, such as singlet-triplet (S-T) gaps or $J$-coupling parameters may often not be resolved within chemical accuracy without the additional inclusion of dynamic correlation effects\cite{NeeseDynamicCASSCF}. The historically most popular and commonly used method to account for post-CI dynamic correlation CASSCF in combination with second-order many-body perturbation theory (CASPT2) suffers from a variety of shortcomings, including poor computational scaling, and convergence issues arising from the fact that the MP2 correction is not variational, often leading to nonphysical lower bounds to the total electronic energy\cite{CASPT2-1, CASPT2-2, CASPT2-3, CASPT2-4}. Consequently, the development of electronic structure methods that account for post-CI dynamic correlation is an area of major research interest and recent developments include algorithms such as quantum Monte-Carlo\cite{QMC1, QMC3, QMC4}, multi-configuration pair-density functional theory (MC-PDFT)\cite{PDFT1, PDFT2, PDFT3, PDFT4, PDFT5}, reduced-density-matrix functional theory (RDMFT)\cite{RDMFT1, RDMFT2, RDMFT3, Piris2021, Schilling2021, Schmidt2021}, incremental FCI (iFCI)\cite{iFCI1, iFCI2, iFCI3} or CASCI in combination with the anti-Hermitian contracted Schr\"odinger equation (ACSE)\cite{MCACSE1, MCACSE2} as well as related methods that use cumulant reconstruction\cite{schwinger, cumulant} to solve a contracted Schr{\"o}dinger equation\cite{CSE1, CSE2, ACSE} for dynamic correlation\cite{Chan2007, Li2021}. \\

While FCI yields the exact electronic energy in a chosen basis set and hence is invariant to the molecular orbital (MO) basis, it remains out of reach for system larger than 16 electron in 16 orbitals due to exponential computational scaling. As other ab-initio electronic structure methods that aim to resolve the total electronic correlation energy tend to rely on some approximation to truncate the exact Hamiltonian, they exhibit a dependence on the chosen MO basis. Recent research has been performed in the areas of CC and MP theories with the aim of improving their predictive properties via the use of improved molecular orbitals, rather than the commonly used HF reference\cite{CCSDT-BLYPorbs, CCSDT-orbs, CCSDTorbs-dis, OOCC, OOCC2}. This includes the implementation of orbital-optimized variants of CC and second-order MP2 (OOMP2), which, while yielding improved results over the HF-reference based implementations, suffer from increased computational scaling, and in the case of OOMP2 three major failures, namely divergence for small MO energy gaps, artificial symmetry restoration and loss of Coulson-Fischer points\cite{OOMP2, OOMP2-2, OOMP2-3}. A contrary approach to the orbital-optimization problem has recently been undertaken by Head-Gordon and coworkers, who demonstrate significant improvements in the prediction of chemical properties in MP3 via the use of OOMP2 and DFT orbitals\cite{MP3-MHG, MP3-MHG-orbs}, and in the calculation of vibrational frequencies with CCSD(T) with the use of DFT orbitals\cite{CCSDT-MHG}. \textcolor{black}{Additionally, natural orbitals obtained with MR-CI-SD calculations performed after initial CASSCF optimization may provide improved orbitals for the recovery of additional correlation energy.\cite{Bytautas2005, Bytautas2007}} \\

While research has been undertaken to shine light on the orbital dependence in single-reference methods aimed at resolving dynamic correlation, work aiming at resolving this dependence in multi-reference and post-multi-reference dynamic correlation calculations has been limited\cite{NOsCAS,CASCIOrbs, CASCIOrbs2, MRPTOrbs, CISNOs, TDDFTNOs, NOsBest, DMRGNOs} and common implementations of electronic structure methods aiming to resolve the total correlation energy such as QMC, CASPT2 or MC-PDFT, tend to rely on CASSCF optimized orbitals as their reference. But are orbitals that are optimized to include multi-reference correlation necessarily the best to account for the total correlation or is the restriction of the orbital optimization to an active space representing a small subset of the total molecular orbitals hindering the capture of the complete electronic structure? Specifically, would CASSCF orbitals necessarily provide the best initial guess for the orbitals in a post-CASSCF all-electron correlation SCF method? \\

In this article we aim to resolve the orbital dependence of CI and post-CI dynamic correlation calculations by using molecular orbitals obtained from KS-DFT, HF, MP2 and CCSD as reference orbitals in CI calculations, which are then used to seed the anti-Hermitian contracted Schr\"odinger equation (ACSE) to resolve the dynamic correlation. Orbitals obtained from KS-DFT have previously been demonstrated to be more suitable for the construction of electronic states in configuration interaction (CI) calculations compared to HF orbitals\cite{CI-KSorbs} and may provide a viable, cost-saving alternative to CASSCF optimization in the quest to resolve the electronic properties of strongly correlated molecules and materials. We apply the CASCI/ACSE algorithm seeded with the various molecular orbitals from the surveyed single-reference methods to three distinct chemical problems dominated by strong correlation effects, namely the dissociation of N$_2$, the prediction of S-T gaps in a benchmark set of biradicals, and the calculation of the energetic barrier of the isomerization reaction of bicyclobutane to gauche-1,3-butadiene via both the conrotatory and disrotatory transition states. \\

\section{Computational Details}
To investigate the orbital dependence of the static and dynamic parts of the total electronic correlation energy, molecular orbitals were obtained via self-consistent field (SCF) calculations using various popular single-reference, ab-initio methods, \textcolor{black}{as well as CASSCF}. These methods include Hartree Fock (HF), CASSCF, variational 2-RDM CASSCF (V2RDM)\cite{V2RDM}, DFT\cite{KS1}, as well as, MP2 and CCSD, in which case the natural orbitals are investigated. For the DFT calculations, functionals representing the various rungs of Jacobs-Ladder of functional development were chosen, namely simple LDA\cite{LDA}, and the popular functionals PBE\cite{PBE1, PBE2}, BLYP\cite{BLYP1, BLYP2, BLYP3}, B3LYP\cite{B3LYP}, M062X\cite{M062X}, $\omega$B97XD\cite{wB97XD}, MN15\cite{MN15}. Orbitals from these initial SCF calculations were then used to perform a minimal active space complete active space configuration interaction (CASCI) calculation using the V2RDM method with DQGT conditions (V2-T)\cite{V2RDM, Nrep}, obtaining the multi-reference correlation energy in the initial orbitals, as well as the strongly correlated 1- and 2-electron reduced density matrices (RDMs). \\

We then generate the 1- and 2-electron integrals, namely ${}^1K$ containing the kinetic and nuclear attraction integrals and ${}^2V$ containing the electron-electron repulsion integrals, from the molecular orbitals obtained with the selected single-reference method.
These serve as the basis for the ACSE calculations, which is used to calculated the dynamic, post-CI correlation in the given molecular orbital basis. The ACSE arises from the fact that fermions interact pairwise and hence the $N$-electron Schr\"odinger equation may be projected onto the space of only two-electron transitions yielding the contracted Schr\"odinger equation (CSE)\cite{CSE1, CSE2, ACSE}:
\begin{equation}
    \bra{\Psi}\hat{a}^{\dagger}_i \hat{a}^{\dagger}_j \hat{a}_l  \hat{a}_k  \hat{H}\ket{\Psi} = E\; {}^2D^{i,j}_{k,l} \,,
\end{equation}
where $\hat{H}$ is the Hamiltonian operator
\begin{equation}
    \hat{H} = \sum_{ij} {}^1K^i_j \hat{a}^{\dagger}_i \hat{a}_j + \sum_{ijkl} {}^2V^{i,j}_{k,l} \hat{a}^{\dagger}_i \hat{a}^{\dagger}_j \hat{a}_l \hat{a}_k \,,
\end{equation}
\textcolor{black}{
and ${}^2D^{i,j}_{k,l}$ is the 2-RDM:
\begin{equation}
    {}^2D^{i,j}_{k,l} = \braket{\Psi|a_i^\dagger a_j^\dagger a_l a_k|\Psi}  \,.
\end{equation}}
The CSE can be separated into its Hermitian and anti-Hermitian parts, and selection of only the anti-Hermitian part yields the ACSE:
\begin{equation}
    \bra{\Psi}[\hat{a}^{\dagger}_i \hat{a}^{\dagger}_j \hat{a}_l  \hat{a}_k ,\hat{H}]\ket{\Psi} = 0 \,,
\end{equation}
where the square brackets indicate the commutator. Unlike the Hermitian part of the CSE, which depends on the 2-, 3- and 4-RDMs, the highest order terms in the ACSE, which is expanded in more detail in ref\cite{ACSEexpansion}, depend on only the 2- and 3-RDMs. Furthermore, this dependence may be resolved by using an cumulant reconstruction in terms of the 2-RDM\cite{schwinger, cumulant}:
\begin{equation}
    {}^3D^{i,j,k}_{q,s,t} \approx {}^1D^{i}_{q} \wedge {}^1D^{j}_{s} \wedge {}^1D^{k}_{t} + 3 {}^2\Delta^{i,j}_{q,s} \wedge {}^1D^{k}_{t} \,,
\end{equation}
where
\begin{equation}
    {}^2\Delta^{i,j}_{q,s} = {}^2D^{i,j}_{q,s} - {}^1D^{i}_{q} \wedge {}^1D^{j}_{s} \,,
\end{equation}
and $\wedge$ denotes the antisymmetric Grassmann wedge product, which is defined as:
\begin{equation}
    {}^1D^{i}_{k} \wedge {}^1D^{j}_{l} = \frac{1}{2}({}^1D^{i}_{k}{}^1D^{j}_{l} - {}^1D^{i}_{l} {}^1D^{j}_{k}) \,.
\end{equation}
As the 3-RDM terms appear only in the perturbative ${}^2V$ part of the Hamiltonian of the ACSE, this approximate reconstruction of ${}^3D$ neglects the cumulant 3-RDM part of the expansion, setting ${}^3\Delta^{ijk}_{qst}$ to be zero. \\

Using electron integrals and initial guess 1- and 2-RDMs obtained from a lower-level electronic structure calculation of choice, we solve the ACSE via a system of differential equations\cite{ACSEextrap}:
\begin{equation}
\begin{aligned}
    E(\lambda + \epsilon) &= \bra{\Psi(\lambda)}e^{-\epsilon S(\lambda)} \hat{H} e^{\epsilon S(\lambda)}\ket{\Psi(\lambda)} \\
    &= E(\lambda) + \epsilon \bra{\Psi(\lambda)}[\hat{H}, \hat{S}(\lambda)]\ket{\Psi(\lambda)} + O(\epsilon^2) \,,
\end{aligned}
\end{equation}
\begin{equation}
    \frac{dE}{d\lambda} = \bra{\Psi(\lambda)}[\hat{H}, \hat{S}(\lambda)]\ket{\Psi(\lambda)} \,,
\end{equation}
\begin{equation}
    \frac{d{}^2D^{i,j}_{k,l}}{d\lambda} = \bra{\Psi(\lambda)}[\hat{a}^{\dagger}_i \hat{a}^{\dagger}_j \hat{a}_l  \hat{a}_k, \hat{S}(\lambda)]\ket{\Psi(\lambda)} \,,
\end{equation}
where the operator $\hat{S}$ is defined as:
\begin{equation}
    \hat{S}(\lambda) = \sum_{ijkl} {}^2S^{i,j}_{k,l}\hat{a}^{\dagger}_i \hat{a}^{\dagger}_j \hat{a}_l \hat{a}_k (\lambda)  \,,
\end{equation}
chosen at each step of $\lambda$ to minimize the energy along the gradient:
\begin{equation}
    {}^2 S^{i,j}_{k,l}(\lambda) = \bra{\Psi(\lambda)}[\hat{a}^{\dagger}_i \hat{a}^{\dagger}_j \hat{a}_l  \hat{a}_k ,\hat{H}]\ket{\Psi(\lambda)} \,.
\end{equation}
The ACSE is propagated in $\lambda$ until either the energy reaches a minimum or the norm of the residual increases. This algorithm is presented in more detail in Refs.~\onlinecite{ACSEextrap, ACSEexpansion}.\\

Seed 1- and 2-RDMs may be obtained from single- or multi-reference electronic structure calculations, minimizing the total electronic energy in the chosen orbital basis of the electron integrals. When provided with a single-reference guess, such as one obtained from a HF calculation, the ACSE has been demonstrated to yield total electronic energies of comparable accuracy to those from CCSD(T)\cite{ACSECC, ACSE, Alcoba2021}. However, ACSE calculations may also be seeded with initial RDMs from a multi-reference electronic structure calculation, such as CASSCF or CASCI, which yields accurate results even when strong correlation plays a major role\cite{MCACSE1, MCACSE2}. In this case, the ACSE resolves the dynamic correlation on top of the static correlation recovered by the seed RDMs within the orbital basis obtained from the multi-reference calculation. Results have been demonstrated to outperform commonly used CASPT2\cite{ACSEbut, ACSEshift, ACSEES}, and provide comparable accuracy to MC-PDFT and AF-QMC\cite{SA-ACSE}. All calculations were performed with the Quantum Chemistry Package\cite{QCP} as implemented in Maple\cite{maple}. \\

\section{Results}
\subsection{Nitrogen Dissociation}
To investigate the orbital dependence of the single- and multi-reference parts of the electronic correlation energy, we first consider the dissociation of N$_2$. Dissociation of N$_2$ into its two constituent nitrogen atoms provides a classic case of transition from a system dominated by single-reference correlation---N$_2$ near the equilibrium bond length---to a system dominated by multi-reference correlation as the N-N bond is stretched and the natural occupation numbers (NON) in the [6,6] active space formed by the 6 nitrogen-p based orbitals become more and more fractional until they reach full degeneracy in the dissociated regime. \\

\begin{figure*}
    \centering
    \centering
    \begin{minipage}[b]{0.49\textwidth}
        \centering
        \includegraphics[scale=0.23]{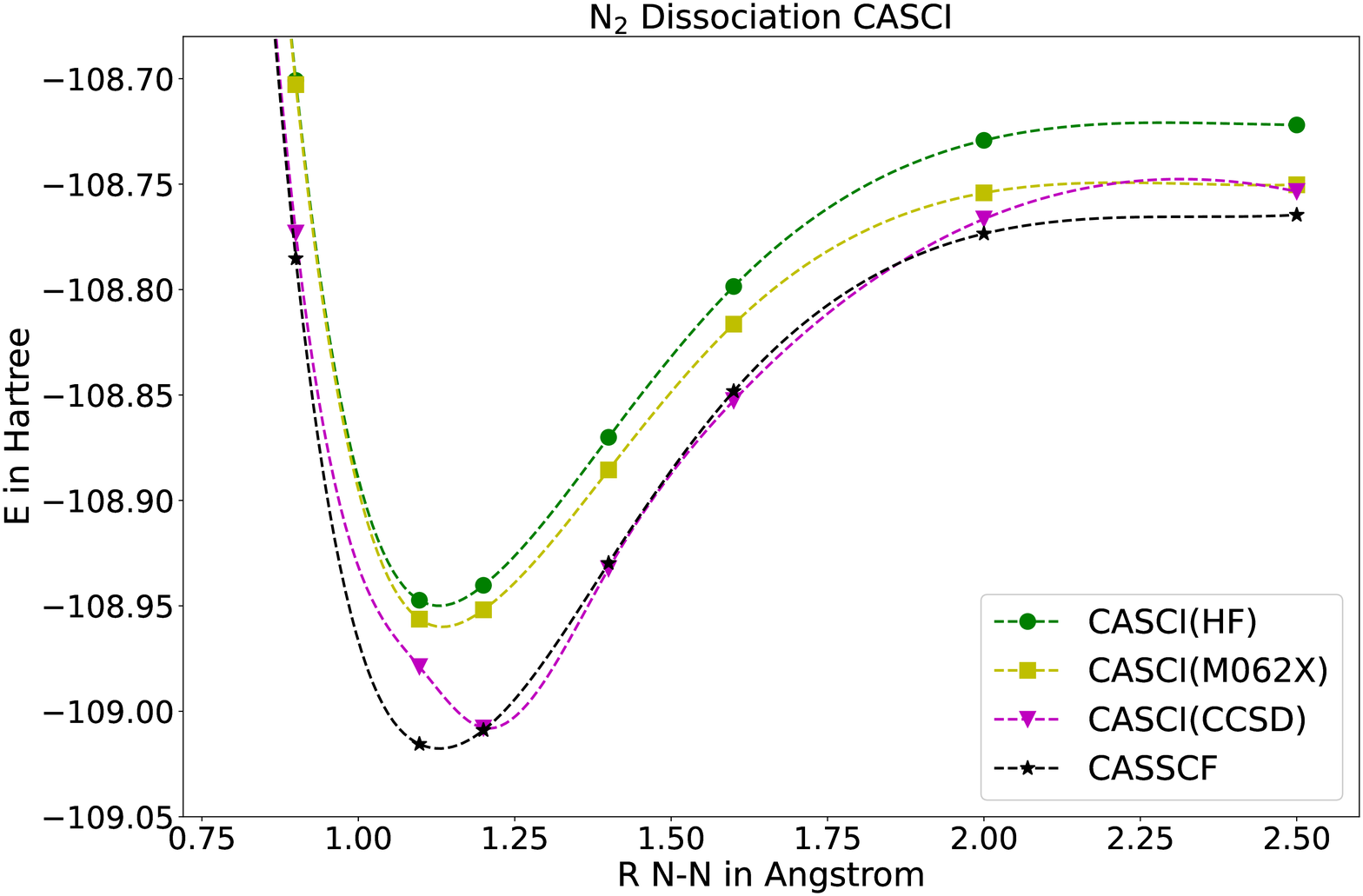}\\
    \end{minipage}
    \begin{minipage}[b]{0.49\textwidth}
        \centering
        \includegraphics[scale=0.23]{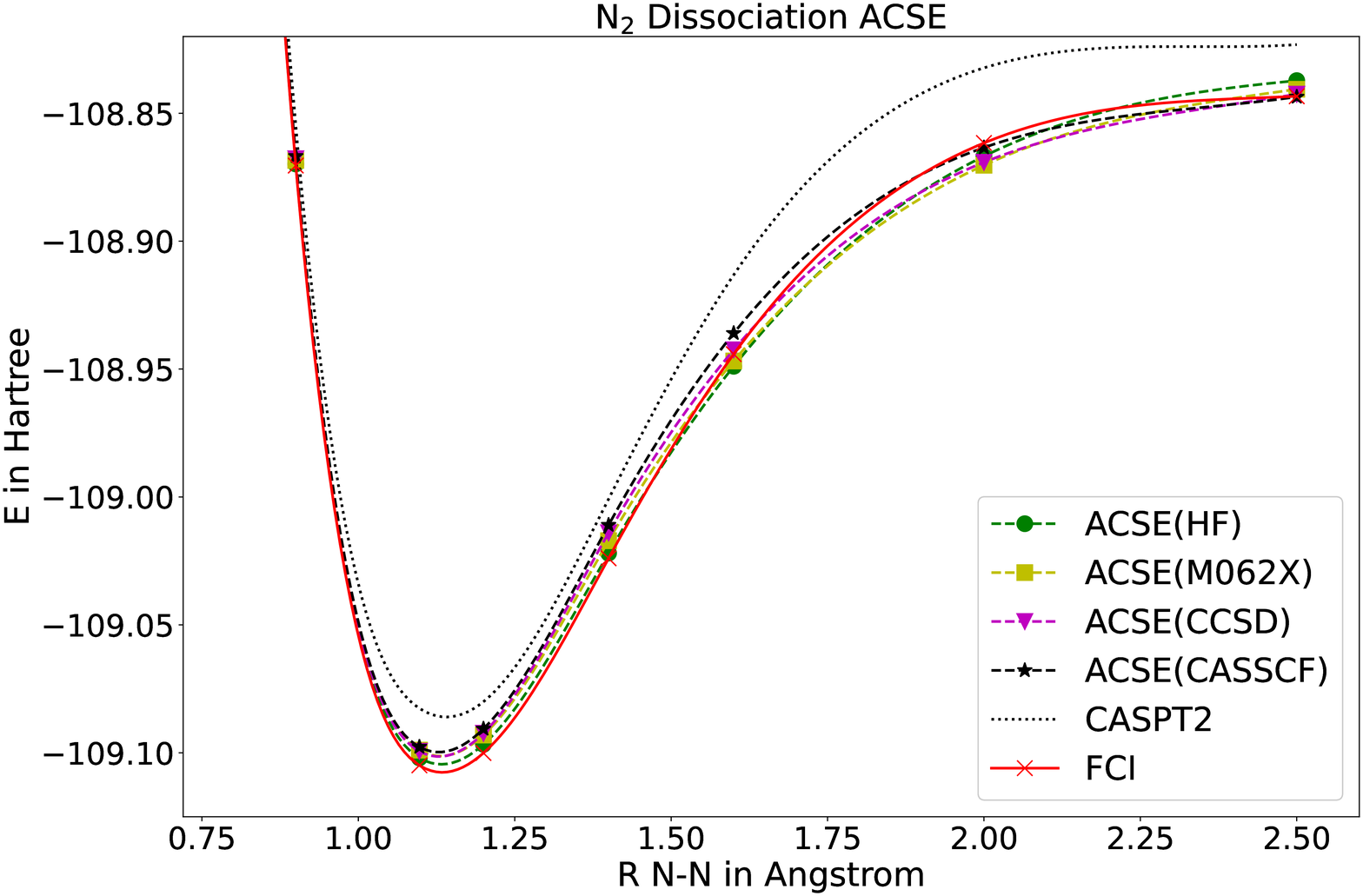}\\
    \end{minipage}
    \caption{Dissociation curves of N$2$ for (left): [6,6] CASCI calculations; (right): ACSE calculations. All calculations were performed with a 6-31G basis set and the ACSE was seeded with the 2-RDM from the [6,6] CASCI calculations. Curves were constructed from eight points along the N-N dissociation coordinate, [0.8, 0.9, 1.0976, 1.2, 1.4, 1.6, 2.0, 2.5] in \AA.\textcolor{black}{We note that the CASCI(CCSD) results yield a lower bound to the CASSCF energy for the $R=1.4$ and $R=1.6$ points. This is the result from minor violation of the full $N$-representability conditions of the V2RDM based CASCI procedure with the DQG and T$_2$ $N$-presentability conditions, which yield a lower bound to the true ground-state energy. For these bond lengths CCSD natural orbitals provide an essential identical solution to CASSCF.}}
    \label{fig:N2dis}
\end{figure*}

For our calculations we consider eight data points of N-N bond lengths, R = [0.8, 0.9, 1.0976, 1.2, 1.4, 1.6, 2.0, 2.5] in \AA. Calculations use the relatively small 6-31G basis\cite{631g} in order to allow comparison to full CI (FCI) data. Once orbitals are obtained via a chosen single-reference method, CASCI calculations in that MO basis set with [6,6] active spaces are carried out to recover the multi-reference correlation in the nitrogen-p-based orbitals. Figure~\ref{fig:N2dis} shows the dissociation curves obtained from the molecular orbitals of a few select methods, with the left panel showing the CASCI results, and the right panel displaying the ACSE results, as well as, the FCI curve. Furthermore, results of the go-to method for the inclusion of post-multi-reference calculations, CASPT2, are also displayed. It is evident that the recovery of the multi-reference correlation is strongly orbital dependent, with large variations in the CASCI curves across the various methods used for orbital optimization. Differences arise not only in terms of the correlation energy recovered across the dissociation coordinate, i.e. in the form of a vertical shift of the curve, but also in its general line shape. This change in line shape is particularly evident in the cases of the HF and CCSD orbitals, which differ significantly from those obtained via DFT. Furthermore, inspection of the ACSE curves shows that while near-exact in the dissociated regime, CASSCF orbitals provide a larger deviation from the FCI curve than any of the displayed single reference methods in the single-reference regime around the equilibrium N-N bond length. \\

\begin{table*}
    \centering
    \begin{tabular}{cc|ccccccccccccc}
        & & \multicolumn{13}{c}{\textcolor{black}{Molecular Orbitals}} \\
        \textcolor{black}{Method} &  & HF & MP2 & CCSD  & CASSCF & CASPT2 & V2-T & LDA & PBE & BLYP & B3LYP & M062X & wB97XD & MN15 \\
        \hline
        \multirow{5}{*}{REF} & MAE & 225.85 & 70.41 & 33.26 &  & 13.76 &  & 947.85 & 131.34 & 201.31 & 161.51 & 157.01 & 160.39 & 115.91 \\
        & MSE & 225.85 & -67.80 & 32.19 &  & 13.76 &  &  947.85 & -131.33 & -201.31 & -161.51 & -153.36 & -160.39 & -104.76 \\
        & $\Delta_{EQM}$ & 148.81 & -0.95 & 5.91 & & 13.99 & & 919.27 & -156.14 & -222.90 & -196.24 & -202.00 & -203.58 & -147.94 \\
        & $\Delta_{DIS}$ & 462.43 & -328.34 & 206.37 &  & 12.68 & & 1041.02 & -31.83 & -108.37 & -33.82 & 14.53 & -6.76 & 44.41 \\
        & $\Delta E_{DIS}$ & 313.62 & -327.38 & 200.45 & & -1.31 & & 121.75 & 124.32 & 114.53 & 162.42 & 216.53 & 196.82 & 192.35 \\
        \hline
        \multirow{5}{*}{CAS} & MAE & 94.03 & 63.73 & 60.82 & 55.49 &  & 54.41 & 84.56 & 82.99 & 83.17 & 83.59 & 85.30 & 84.23 & 85.37 \\
        & MSE & 94.03 & 63.73 & 60.82 & 55.49 &  & 54.41 & 84.56 & 82.99 & 83.17 & 83.59 & 85.30 & 84.23 &  85.37 \\
        & $\Delta_{EQM}$ & 98.88 & 59.11 & 109.01 & 56.09 & & 55.47 & 92.54 & 91.23 & 91.34 & 91.90 &  93.26 & 92.21 & 93.54 \\
        & $\Delta_{DIS}$ & 76.19 & 69.50 & 48.21 & 49.41 & & 49.33 & 54.37 & 52.38 & 52.26 & 53.84 & 58.40 & 55.89 & 55.67 \\
        & $\Delta E_{DIS}$ & -22.69 & 10.38 & -114.82 & -6.68 &  & -6.13 & -38.18 & -38.85 & -39.08 & -38.06 & -34.87 & -36.32 & -37.87 \\
        \hline
        \multirow{5}{*}{ACSE} & MAE & 2.00 & 3.05 & 2.96 & 3.61 &  & 3.02 & 4.18 & 4.44 & 4.82 & 3.80 & 2.93 & 3.48 & 4.40 \\
        & MSE & 0.44 & 3.03 & 1.77 & 3.30 &  & 1.79 & 0.98 & 3.16 & 3.57 & 2.59 & 1.14 & 2.24 & 2.69 \\
        & $\Delta_{EQM}$ & 1.84 & 3.12 & 3.45 & 4.52 &  &  3.97 & 5.02 &  6.62 & 6.94 & 5.46 & 3.73 & 5.11 & 6.46 \\
        & $\Delta_{DIS}$ & 3.84 & 5.24 & 0.51 & -0.14 &  & -0.45 & -2.73 & 0.43 & 0.61 & 1.38 & 1.73 &  1.54 & -0.67\\
        & $\Delta E_{DIS}$ & 2.01 & 2.12 & -2.94 & -4.66 &  & -4.43 & -7.76 & -6.18 & -6.33 & -4.08 & -2.00 & -3.57 & -7.13 \\
    \end{tabular}
    \caption{Results for the various reference calculations used for the orbital optimization, as well as [6,6] CASCI, and CASCI/ACSE calculations \textcolor{black}{for the dissociation of $N_2$, in kcal/mol}. All calculations were performed with the 6-31G basis set. Errors are relative to the FCI energies and MSE and MAE are calculated over the eight distinct points along the dissociation coordinate; $\Delta_{eqm}$ and $\Delta_{dis}$ are the errors at $R=1.0976$ and $R=2.5$, respectively; and $\Delta E_{DIS}$ is the error in the dissociation energy, $E_{DIS} = E_{R=1.0976} - E_{R=2.5}$, with respect to FCI.}
    \label{tab:N2data}
\end{table*}

To allow a more in-depth analysis of the optimality of the molecular orbitals from a chosen method in accounting for the different parts of the total electronic correlation energy, we consider the errors of the energies obtained via CASCI relative to the CASSCF results, and the energies obtained via CASCI/ACSE relative to the FCI results. Table \ref{tab:N2data} shows the mean absolute error (MAE) and mean signed error (MSE) versus FCI over the eight N-N bond lengths, as well as, the maximum and minimum errors. The initial results from the orbital optimization calculations split as expected with the wave-function based methods providing upper bounds to the FCI energy and DFT yielding lower bounds. The largest positive deviation from FCI \textcolor{black}{data} results from HF with a MSE of 225.85 kcal/mol, while the largest negative deviation \textcolor{black}{with} DFT is obtained with the BLYP functional at an MSE of -201.31 kcal/mol. CCSD yields the best results with an MSE of \textcolor{black}{33.19} kcal/mol, outperforming MP2, which as expected results in unphysical behavior in the dissociated regime and hence large negative deviations from FCI. Use of simple LDA gives rise to unphysical electronic energies with a MAE of 947.85 kcal/mol. \\

If we consider the contribution of the multi-reference correlation to the total electronic energy, CASSCF calculations using the minimal [6,6] active space give a MSE of 55.49 kcal/mol vs FCI, and on average the correlation recovered in the [6,6] active space accounts for 69.4\% of the total correlation energy in the 6-31G basis set across the 8 dissociation points. The CASSCF calculation provides the benchmark result to assess the ability of a chosen method's orbitals to account for multi-reference correlation. As a CASSCF calculation uses orbital rotations to minimize the total energy as a functional of the CI energy in the active space, it yields the variational minimum to the multi-reference correlation that may be recovered in the chosen [6,6] active space and 6-31G basis set, and all CI calculations performed on different sets of orbitals yield upper bounds to this energy. Of the surveyed single-reference methods, the NOs from a CCSD calculation provide the orbitals that best account for static correlation across the N$_2$ dissociation space and yield the lowest CASCI energy, with a MAE of 60.82 kcal/mol, followed by MP2, which even though giving rise to nonphysically low energies in the dissociated regime gives natural molecular orbitals that recover a correct CASCI picture with a MAE of 63.73 kcal/mol. DFT yields better multi-reference-ready orbitals than HF, with variation across the different surveyed DFT functionals being relatively small, ranging from the most correlated orbitals at PBE, MAE of 82.99 kcal/mol, to MN15 at MAE of 85.37 kcal/mol, recovering on average roughly 50\% of the total correlation energy. \\

Application of the ACSE following the converged CASCI calculation in the chosen molecular orbital basis resolves dynamic correlation with accuracy comparable to CCSD with perturbative triple excitations [CCSD(T)], in addition to the multi-reference correlation recovered by the CI. Across all surveyed orbitals the CASCI/ACSE recovers an average of between 97.2\% and 99.7\% of the total correlation energy, displaying relatively minor dependence on the chosen molecular orbital basis compared to CI and single-reference calculations. Surprisingly, while providing the best multi-reference orbitals, CASSCF does not provide the most optimal orbitals to resolve the total correlation energy, recovering on average 98\% of the total correlation energy with a MAE of 3.61 kcal/mol. As such, CASSCF orbitals are outperformed by orbitals obtained with HF, MP2, CCSD, M062X, and wB97XD calculations. Simple HF provides the most optimal orbitals to account for both strong and post-CI dynamic correlation, recovering an average of 99.7\% of the FCI energy for MAE of 2.00 kcal/mol. Results from the different DFT functionals vary from an MAE of 2.93 kcal/mol in M06-2X to 4.82 kcal/mol in BLYP. Even in the case of the worst performing DFT functional, BLYP, the MAE increases by only 1.21 kcal/mol over CASSCF. The fact that MSEs are smaller than the MAEs result from the fact that the ACSE may yield a slightly lower bound to the FCI energy in the dissociated regime. ACSE calculations significantly outperform CASPT2, which only recovers an average of 93.0\% of the total correlation energy, for MAE of 13.76 kcal/mol.  \\

Lastly, after considering recovery of the full FCI dissociation curve, we also consider the errors in the reproduction of the FCI dissociation energy, $\Delta E_{DIS} = E_{DIS,method} - E_{DIS,FCI}$, where $E_{DIS} = E_{R=1.0976} - E_{R=2.5}$. This provides a benchmark for orbitals obtained with a certain method to accurately recover the total energy in the dissociated multi-reference regime and the single-reference regime around the equilibrium bond distance. The observed trends follow those discussed above with large positive errors in the DFT, HF and CCSD reference calculations, which significantly overestimate the bonding energy as they break down as N$_2$ is dissociated, while MP2 diverges to large negative energies. Considering the CASCI energies, all calculation underestimate $E_{DIS}$, as correlation is more accurately captured at long bond distances. Now, accounting for mostly static correlation with [6,6] CASCI calculations, CASSCF, and the V2RDM implementation of CASSCF yields the optimal orbitals, followed by MP2 NOs, then HF and finally the various DFT functionals which display only minor variations. CCSD presents an outlier as the CASCI calculation with CCSD NOs at equilibrium suffered from convergence issues. Inspection of the individual errors at the equilibrium and dissociated geometry, $\Delta_{EQM}$ and $\Delta_{DIS}$, respectively, reveals the HF result to arise from a favorable cancellation of error, with the energy lying high above the CASSCF and FCI references. Additionally, as indicated by $\Delta_{EQM}$ and $\Delta_{DIS}$, the single reference orbitals tend to yield significantly lower errors in the multi-reference, dissociated regime than the dynamically correlated regime around the equilibrium bond length. \\

Finally, accounting for all-electron correlation with the CASCI/ACSE method, yields $\Delta E_{DIS}$s within 10 kcal/mol of the FCI result for all surveyed orbitals. Errors range in magnitude from a minimum of 2.00 kcal/mol with MOs from the M062X DFT functional and 2.01 kcal/mol with HF orbitals to 7.76 kcal/mol with LDA DFT MOs. Only orbitals from HF and MP2 result in an overestimation of $E_{DIS}$, while all others yield a negative deviation from FCI. Interestingly, CASSCF does not yield the optimal orbitals to resolve the electronic energy accurately in both the equilibrium and dissociated regimes, with deviations from FCI of -4.66 and -4.43 kcal/mol for its wavefunction and V2RDM implementations, respectively. Indeed, separate consideration of the errors at equilibrium and dissociation bond lengths reveals a relatively large $\Delta_{eqm}$ of 4.52 kcal/mol for the CASSCF orbitals, compared to only 1.84 kcal/mol, 3.12 kcal/mol and 3.45 kcal/mol for those obtained with single-reference methods HF, MP2 and CCSD, respectively. Interestingly, for the DFT MOs, $\Delta_{eqm}$ is larger that obtained with CASSCF MOs for all functionals but M06-2X, with generally lower errors in the multi-reference, dissociation regime. The result is that CASSCF/ACSE all-electron correlation calculations with initial CASSCF optimization of the orbitals for the dissociation energy of N$_2$, where energy differences between a strongly correlated and a dynamically correlated solution are computed, are outperformed by simple CASCI/ACSE calculations using orbitals obtained from HF, MP2, CCSD, as well as the DFT functionals B3LYP, M062X, and wB97XD. \\

\subsection{Singlet Triplet Gaps of Main Group Biradicaloids}
Biradicals play important roles in a wide range of chemical processes as transition states and intermediates, formed during the breaking and forming of chemical bonds, as well as in the development of new single-molecule magnets and materials for spintronics or molecular qubits\cite{Diradicals, Diradicals2, Diradicals3}. Their accurate theoretical description continues to pose a challenge to current developments in electronic structure theory due to the multi-reference character of their open-shell singlet states, requiring the use of methods that account for both static and dynamic correlation to resolve their electronic properties. In this section we investigate the orbital dependence of the singlet-triplet (S-T) gaps for a benchmark set of small main group biradicals OH$^+$, NH, NF, O$_2$, NH$_2^+$, CH$_2$, PH$_2^+$, and SiH$_2$ as calculated with single-reference, CI and post-CI ACSE methods. Singlet-triplet gaps of biradicaloids are of particular value as they relate to experimentally relevant properties such as the exchange coupling constant $J$ or their properties in photonics. \\

\begin{table*}
    \centering
    \begin{tabular}{cc|ccccccccccc}
       & & \multicolumn{11}{c}{\textcolor{black}{Molecular Orbitals}} \\
       \textcolor{black}{Method} &  & HF & MP2 & CCSD  & V2-T & LDA & PBE & BLYP & B3LYP & M062X & wB97XD & MN15 \\
       \hline
        \multirow{4}{*}{REF} & MAE & 20.3 & 70.7 & 6.5 &  & 21.4 & 11.5 & 7.12 & 7.19 & 6.62 & 7.01 & 4.7 \\
        & MSE  & 20.3 & 70.7 & 6.5 &  & 21.4 &  11.5 & 6.7 & 7.19  & 6.62 & 6.96 & 3.63 \\
        & $\Delta_{\text{max}}$ & 31.1 & 93.5 & 13.1  &  & 35.7 & 22.7 & 16.7 & 17.1 & 13.71 & 16.82 & 10.9 \\
        & $\Delta_{\text{min}}$ & 14.6 & 26.3 & 0.12 &  & 12.4  & 3.8 & 0.03 & 0.05 & 0.77 & 0.10 & 1.08 \\
        \hline
         \multirow{6}{*}{CAS} & MAE & 15.48 & 14.45 & 9.39 & 6.38 & 11.43 & 9.75 & 10.57 & 10.75 & 11.85 & 11.18 & 10.28 \\
         & MSE & 10.88 & -3.16 & -4.18 & -0.56 & 5.16 & 5.37 & 4.06 & 4.95 & 6.01 & 4.88 & 3.96  \\
         & $\Delta_{\text{max}}$ & 19.96 & 45.57 & 45.57 & 16.68 & 25.07 & 17.54 & 26.03 & 23.21 & 23.37 & 25.17 & 25.26 \\
         & $\Delta_{\text{min}}$ & 9.74 & 0.04 & 0.12 & 0.04 & 7.37 & 4.89 & 4.88 & 6.36 & 8.24 & 6.37 & 4.51 \\
         & $\bar{R}_{S}$ & 0.251 & 0.212 & 0.628 &  0.912 & 0.855 & 0.888 & 0.886 & 0.851 & 0.720 & 0.818 & 0.866 \\
         & $\bar{R}_{T}$ & 2.005 & 2.027 & 2.037 &  2.036  & 2.013 & 2.013 & 2.013  & 2.011 & 2.009 & 2.011 & 2.013 \\
         \hline
         \multirow{6}{*}{ACSE} & MAE & 5.44 & 7.36 & 3.64 & 3.16 & 4.63 & 5.22 & 4.37 & 3.99 & 4.07 & 3.99 & 4.48  \\
         & MSE & 2.18 & 12.47 & -0.81 & -1.28 & -3.27 & -1.64 & -2.98 & -2.87 & -1.85 & -2.43 & -3.10  \\
         & $\Delta_{\text{max}}$ & 8.71 & 12.47 & 9.44 & 6.25 & 9.61 & 12.44 & 10.34 & 9.56 & 9.62 & 9.47 & 10.61 \\
         & $\Delta_{\text{min}}$ & 0.06 & 0.05 & 0.36 & 0.36 & 0.14 & 0.51 & 0.42 & 0.21 & 0.06 & 0.45 & 0.58 \\
         & $\bar{R}_{S}$ & 0.533 & 0.432 & 0.835 & 1.102  & 1.067 & 1.116 & 1.109 & 1.084 & 0.968 & 1.055 & 1.084 \\
         & $\bar{R}_{T}$ & 2.189 & 2.199 & 2.197 & 2.200 & 2.166 & 2.182 & 2.180 & 2.180 & 2.183 & 2.183 & 2.172 \\
    \end{tabular}
    \caption{\textcolor{black}{Errors} for the S-T gaps of the set of eight biradicals, OH$^+$, NH, NF, O$_2$, NH$_2^+$, CH$_2$, PH$_2^+$, and SiH$_2$. All calculations were carried out using the cc-pVTZ basis set, and CASCI and ACSE calculations use a [4,4] active space. MAE and MSE, and maximum and minimum absolute errors, $\Delta_{\text{max}}$ and $\Delta_{\text{min}}$, are calculated relative to experimental reference values obtained from references\cite{HuberConst, NH2, CH2, PH2, SiH2} \textcolor{black}{and given in kcal/mol}. $\bar{R}_{S}$ and $\bar{R}_{T}$ denote the average distance of the NON from the HF solution for the singlet and triplet states, respectively.}
    \label{tab:STgaps}
\end{table*}

We perform calculations on the singlet and triplet states with the cc-pVTZ basis set\cite{basis1, basis2}, using a minimal [4,4] active space chosen around the HOMO and LUMO for the CI calculations, allowing for the inclusion of non-trivial correlation in the triplet states and assessing the suitability of the single-reference orbital to yield a suitable CAS guess. The ACSE calculations use a spin-averaged implementation, which has recently been demonstrated to yield highly accurate singlet-triplet gaps for this benchmarking set with converged CASSCF wave functions\cite{SA-ACSE}. The data are presented in Table \ref{tab:STgaps}. Errors are calculated with respect to the experimental reference values. As expected, none of the single-reference methods yield accurate results, failing to capture the strong correlation of the biradical singlets. MP2 yields extremely inaccurate results with a MAE of 70.7 kcal/mol, while the rest of the surveyed methods range from 6.5 kcal/mol in CCSD (use of perturbative triples correction reduces this to 4.6 kcal/mol) to 21.4 kcal/mol in LDA. HF also yields a large MAE of 20.3 kcal/mol. Across the surveyed DFT functionals we observe a quite significant variation in their ability to calculate the S-T gaps, with MN15 giving the lowest MAE of 4.7 kcal/mol and LDA and PBE giving the largest MAEs of 21.4 kcal/mol and 11.5 kcal/mol, respectively. \\

As with the dissociation of N$_2$, of the surveyed methods CCSD natural orbitals yield the most optimal basis to account for multi-reference character in the CASCI calculations, with a MAE of 9.39 kcal/mol, followed by the various DFT functionals, where variation in the CASCI results is less pronounced than in the single-reference calculations. Furthermore, there is no correlation between the accuracy of the CASCI calculation and the reference DFT calculation with orbitals from the previously best performing MN15 now yielding a MAE of 10.28 kcal/mol while the second-worst performing PBE returns the best CASCI results with a MAE of 9.75 kcal/mol. Apart from the unreliable MP2 calculations, HF orbitals give the largest MAE of 15.6 kcal/mol. No orbital basis from any method comes close in accuracy to the CASSCF calculation, which yields a MAE of 6.38 kcal/mol. While the MSE is close to zero (-0.56 kcal/mol) in CASSCF, it is of significant, positive magnitude in DFT, ranging from 3.96 kcal/mol in MN15 to 6.01 kcal/mol in M06-2X and HF, while it is negative in MP2 and CCSD. As an additional measure to probe the bi-radical character in the solutions, we introduce the average distance of the 1-RDM of the CASCI calculation from a closed-shell single-reference 1-RDM, defined as $\bar{R} = \sum_{ij}{ ||\lambda_{HF,i} - \lambda_i |_{j}/N}$, where $i$ runs over all orbitals of the system, $N$ is the number of species in the set, $j$ runs over all its members, $\lambda_{i}$ denotes the $i$th natural occupation number and $\lambda_{HF,i}$ is 2 if the orbital is occupied and 0 if the orbital is virtual. \\

The inclusion of post-CI dynamic correlation with the spin-averaged ACSE provides significant improvement over single-reference and CASCI results in all cases. While the [4,4] CASSCF optimization does give the best agreement with experiment (MAE of 3.16 kcal/mol), the advantage over the various other orbitals is relatively minor, with MP2 and HF orbitals yielding the largest deviations with MAEs of 7.36 kcal/mol and 5.44 kcal/mol, respectively, and CCSD again yielding the closest agreement with an MAE within 0.5 kcal/mol of CASSCF. The CASCI/ACSE calculations performed with DFT orbitals provide MAEs within 1 kcal/mol of the CASSCF orbitals for all functionals but PBE, which has the largest error at an MAE of 5.22 kcal/mol. Across the surveyed functionals, variation again is minor, meaning while S-T gaps predicted by the individual functionals differ based on empirical fitting, the underlying molecular orbitals obtained in the SCF procedure remain relatively unchanged. Analogous to the N$_2$ dissociation, the $\omega$B97X-D functional again performs well and provides the molecular orbitals best suited to account for the multi-reference and dynamic correlation in the biradicaloid set, yielding a MAE of 3.99 kcal/mol. In the CASCI/ACSE case the sign of the MSE obtained for the various DFT orbitals agrees with CASSCF, and only HF and MP2 orbitals lead to a positive MSE, while CCSD yields a MSE of small negative magnitude. Again considering $\bar{R}$  as a measure for the recovered total correlation, HF and MP2 which yield the largest MAEs also result in the lowest magnitude of $\bar{R}_{S}$, 0.533 and 0.432, respectively. However, surprisingly, $\bar{R}_{S,CCSD} = 0.835$ is significantly lower than  $\bar{R}_{S,CASSCF} = 1.102$, while DFT orbitals, yield $\bar{R}_{S}$ values between 0.968 and 1.116, with there being no correlation between $\bar{R}_{S}$ and MAE. To the contrary, PBE yields greater correlation in the NOs than CASSCF while resulting in the largest MAE of all surveyed functionals. \\

\begin{figure*}
    \centering
    \begin{minipage}[b]{0.49\textwidth}
        \centering
        \includegraphics[scale=0.28]{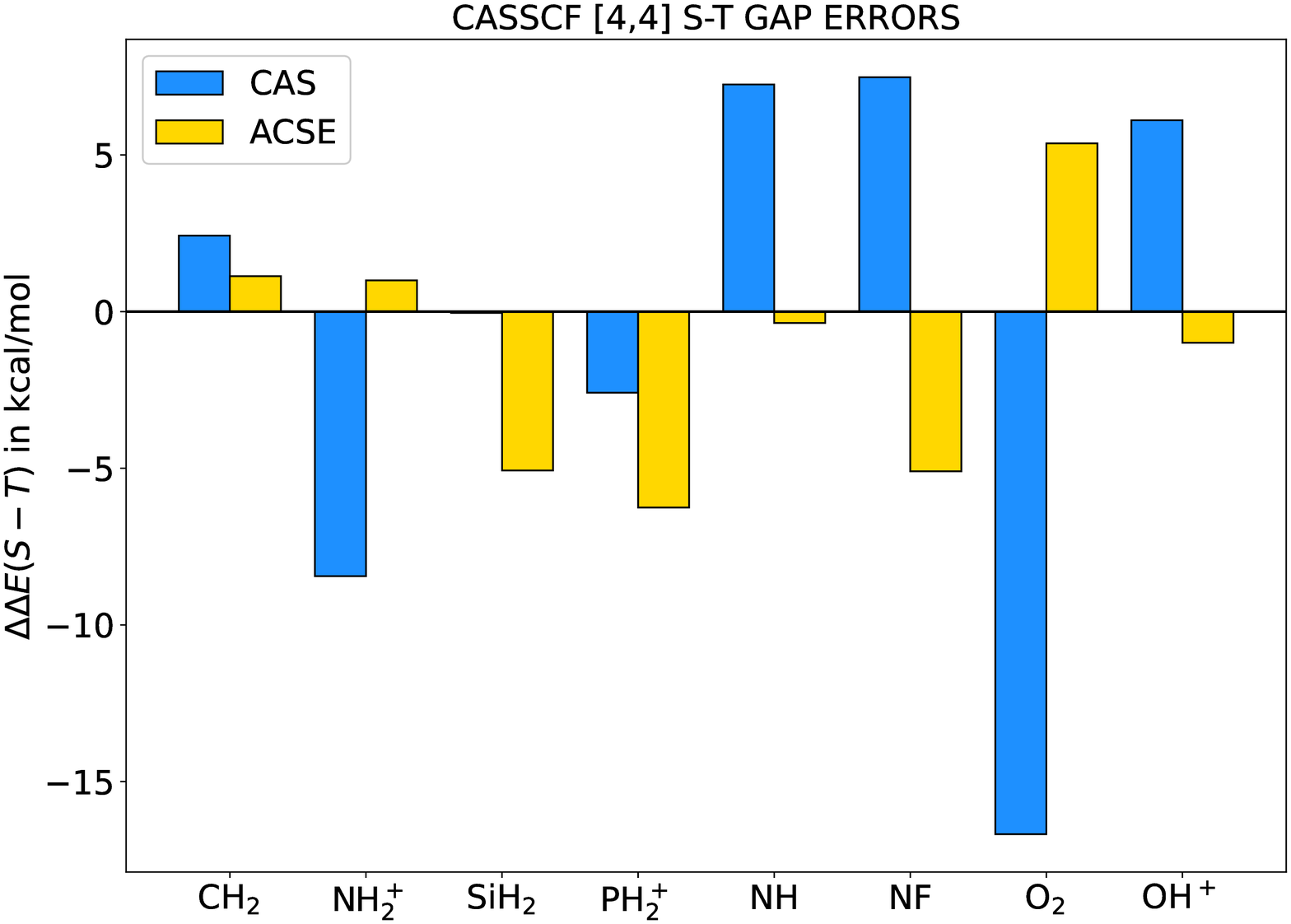}\\
        \includegraphics[scale=0.28]{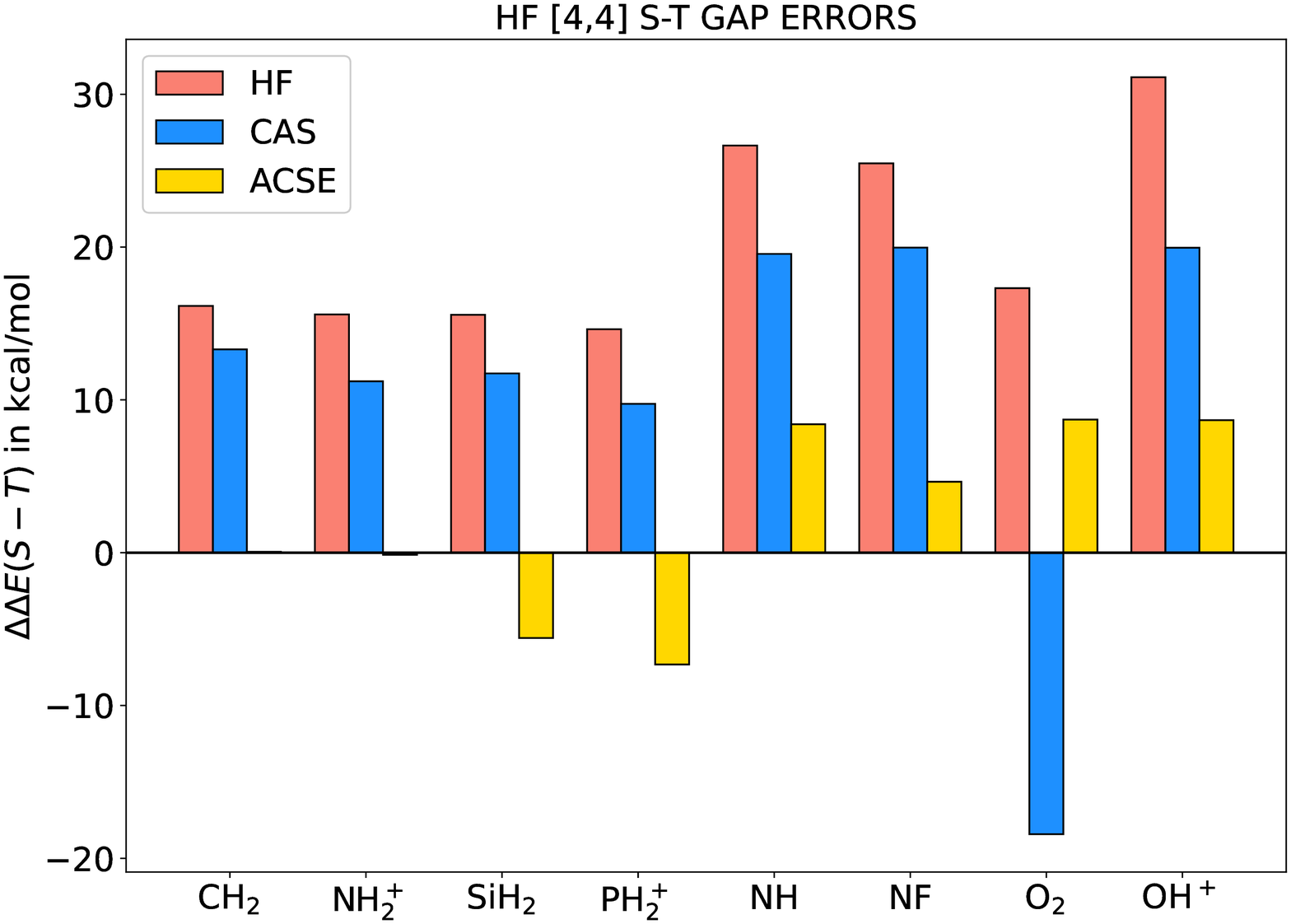}\\
    \end{minipage}
    \begin{minipage}[b]{0.49\textwidth}
        \centering
        \includegraphics[scale=0.28]{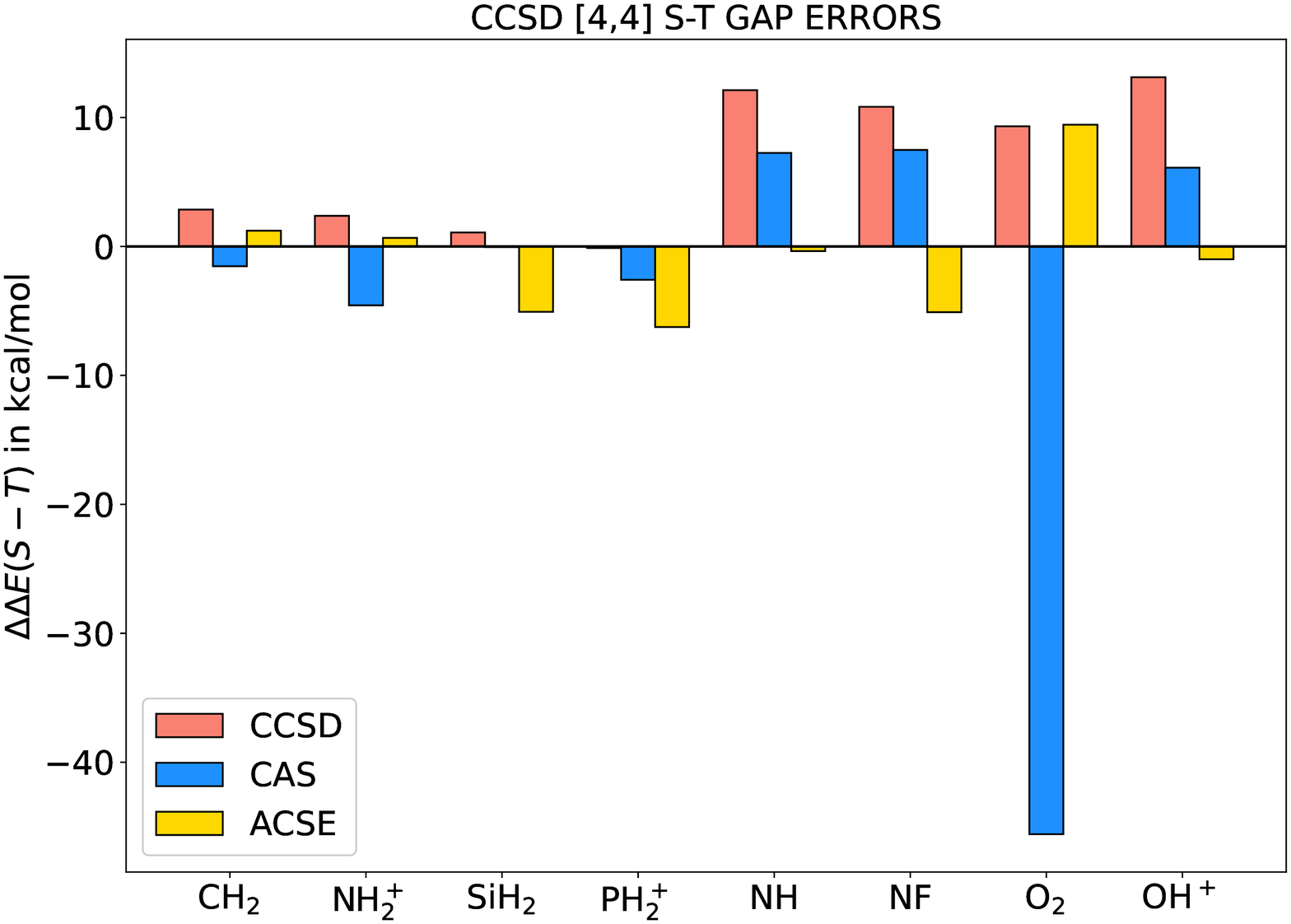}\\
        \includegraphics[scale=0.28]{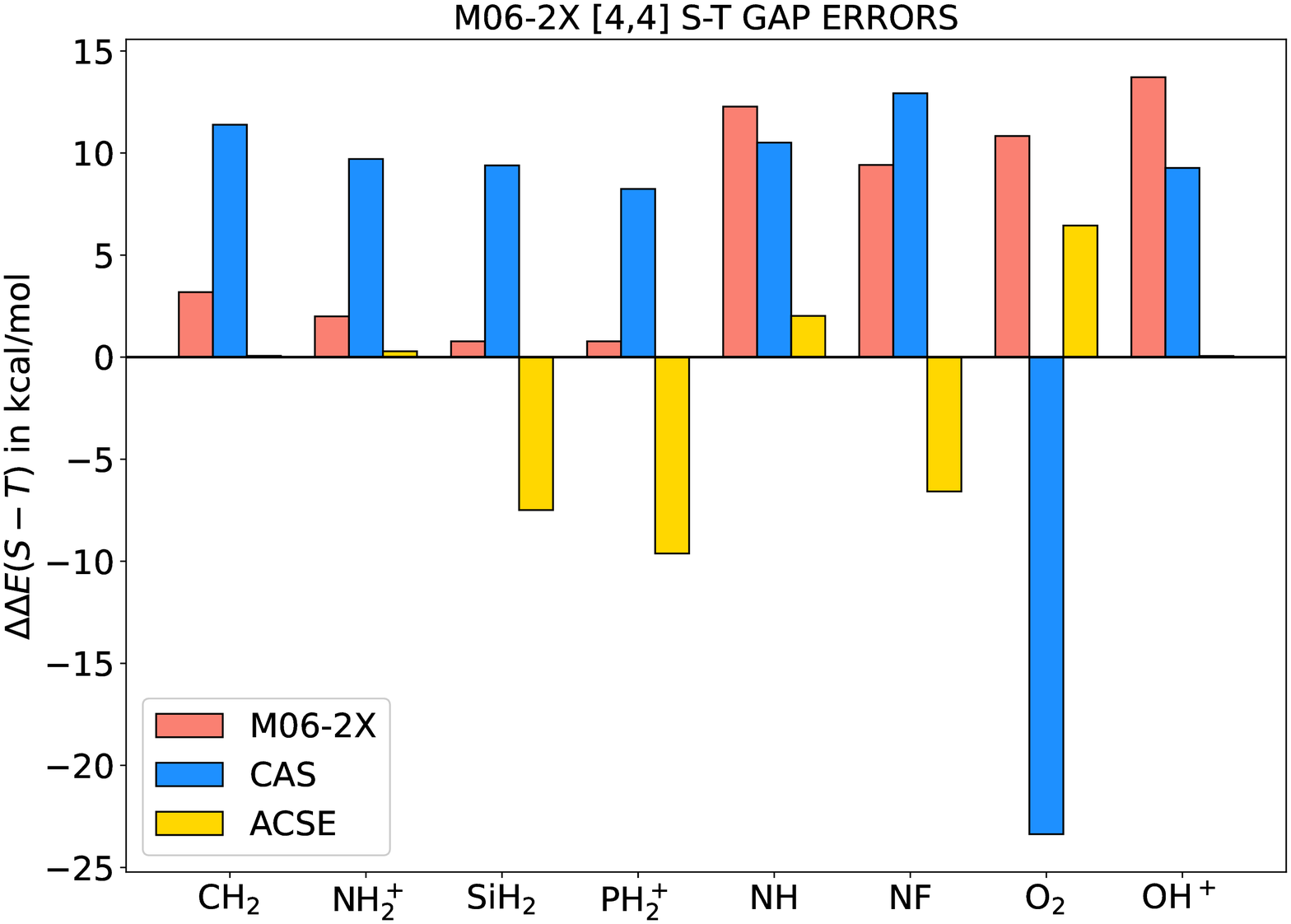}\\
    \end{minipage}
    \caption{\textcolor{black}{Errors} for the S-T gaps of the biradical set resolved for its individual members for four select methods used for the orbital optimization. Bars indicate the errors of the S-T gap with respect to the experimental reference. Orange bars indicate errors of the single-reference calculation, blue bars the CASCI results based on the single-reference orbitals, and yellow bars the CASCI/ACSE result in the respective single-reference orbital basis. Top row: CASSCF (left), CCSD (right); bottom row: HF (left), M06-2X (right). All data were obtained with a cc-pVTZ basis set and [4,4] active spaces for the CASCI and ACSE calculations. }
    \label{fig:STerrors}
\end{figure*}

To resolve the origin of the minor variations across surveyed methods, we plot the deviations from experimental reference values for the individually studied species for CASSCF, HF, CCSD, and M06-2X in Figure~\ref{fig:STerrors}. Inspection of the individual errors in CASCI shows particularly large errors in the calculation of O$_2$, which has previously been demonstrated to provide a challenge to various electronic structure methods, with AFQMC and ACSE calculations requiring CASSCF wave functions with active spaces as large as [10,15] and [14,14] to yield sub 2 kcal/mol accuracy for the calculation of its S-T gap\cite{SA-ACSE, Reichman}. The [4,4] active space successfully resolves its diradical character; however, it significantly overstabilizes the singlet state leading to a large negative deviation from experiment. We observe a particularly strong stabilization of the singlet O$_2$ state in the CASCI with CCSD molecular orbitals, while the remaining orbitals yield results that are in good agreement with CASSCF, leading to the more negative MSE in the CCSD NO basis. The positive MSE in DFT and HF arises from the fact that, while there is agreement with CASSCF in the case of O$_2$, the orbitals in all other species lead a to an overestimation of the singlet triplet gap via a relatively less pronounced stabilization of the singlet. Use of the ACSE to include post-CI correlation leads to a significant reduction in the variation of the errors across the different species and methods, with only HF orbitals showing deviations in significant magnitude from the CASSCF/ACSE results, particularly in the cases of NF and OH$^{+}$.

\begin{table}
    \centering
    \begin{tabular}{cc|cccc}
         & & \multicolumn{4}{c}{\textcolor{black}{Molecular Orbitals}} \\
         & & M06-L & M06 & M06-2X & M06-HF \\
        \textcolor{black}{Method} & \% HF & 0 & 27 & 54 & 100\\
         \hline
         \multirow{4}{*}{REF} & MAE & 8.65 & 6.92 & 6.62 & 4.25 \\
         & MSE & 8.65 & 4.73 & 6.62 & 4.22 \\
         & $\Delta_{\text{max}}$ & 20.62 & 15.76 & 13.71 & 11.34\\
         & $\Delta_{\text{min}}$ & 0.19 & 0.75 & 0.77 & 0.13 \\
         \hline
         \multirow{6}{*}{CAS} & MAE & 10.65 & 10.77 & 11.85 & 14.41\\
         & MSE & 4.40 & 5.10 & 6.01 & 4.55\\
         & $\Delta_{\text{max}}$ & 25.00 & 22.70 & 23.37 & 21.39 \\
         & $\Delta_{\text{min}}$ & 5.45 & 7.06 & 8.24 & 9.37\\
         & $\bar{R}_{S}$ & 0.884 & 0.838 & 0.720 & 0.396 \\
         & $\bar{R}_{T}$ & 2.014 & 2.011 & 2.009 & 2.005 \\
         \hline
         \multirow{6}{*}{ACSE} & MAE & 3.92 & 4.26 & 4.07 & 4.55 \\
         & MSE & -2.52 & -2.95 & -1.85 & -0.27 \\
         & $\Delta_{\text{max}}$ & 8.79 & 10.65 & 9.62 & 9.72\\
         & $\Delta_{\text{min}}$ & 0.03 & 0.25 & 0.06 & 0.52\\
         & $\bar{R}_{S}$ & 1.122 & 1.072 & 0.968 & 0.662 \\
         & $\bar{R}_{T}$ & 2.190 & 2.179 & 2.183 & 2.177 \\
    \end{tabular}
    \caption{\textcolor{black}{Errors} for the biradical set S-T gaps resolved with orbitals from the members of the MN06 suite of functionals. \textcolor{black}{All calculations were carried out using the cc-pVTZ basis set, and CASCI and ACSE calculations use a [4,4] active space. MAE and MSE, and maximum and minimum absolute errors, $\Delta_{\text{max}}$ and $\Delta_{\text{min}}$, are calculated relative to experimental reference values obtained from references\cite{HuberConst, NH2, CH2, PH2, SiH2} and given in kcal/mol. $\bar{R}_{S}$ and $\bar{R}_{T}$ denote the average distance of the NON from the HF solution for the singlet and triplet states, respectively.}}
    \label{tab:M06_exchange_ST}
\end{table}

Lastly, to provide insight into the effect of exact HF exchange in a chosen DFT functional on the molecular orbitals obtained from the SCF procedure, we compare the results from four M06 functionals with varying degrees of HF exchange: M06-L (0\%), M06 (27\%), M06-2X (54\%), and M06-HF (100\%), shown in Table \ref{tab:M06_exchange_ST}. Interestingly, the MAE and MSE of the DFT S-T gaps decreases as the HF exchange contribution to the functional increases. While this is contrary to results from large-scale functional benchmarks, which suggest functionals with larger HF exchange contributions yield worse performance on multi-reference interactions\cite{DFTBenchmark}, the expected trend is observed in the CASCI errors. These consistently increase with the HF exchange fraction, suggesting the inclusion of more exact HF exchange leads to worse multi-reference orbitals in the KS-SCF optimization. Inclusion of post-CI dynamic correlation again results in reduced variation across the functionals, however, showing a trend of increasing MAE with increasing HF exchange in the functional used for the orbital optimization. In fact, the molecular orbitals obtained from a SCF optimization with the 0\% HF exchange containing M06L functional yields the lowest MAE, as well as, maximum and minimum errors across the data set of all surveyed functionals. Additionally, the decrease in the ability of the functional's orbitals to account for multi-reference character of the singlet state is clearly reflected in the trend observed in the $\bar{R}_{S}$ values with a steady decrease from a maximum of 1.122 in M06-L to just 0.662 in M06-HF. \\

While the results obtained from the [4,4] spin-averaged CASSCF/ACSE calculation are comparable to those from CASSCF/MC-PDFT, which yields a MAE of 3.5 kcal/mol\cite{PDFT-MG}, the computationally less expensive CASCI/ACSE calculations based on a DFT orbital optimization yield results in line with various methods reported across the literature, such as (V)FS-PBE (MAE = 4.3 kcal/mol)\cite{VFS}, pp-B3LYP (4.8 kcal/mol)\cite{YangPPrpa}, W2X (3.7 kcal/mol)\cite{PDFT-MG}, or tPBE/MC-PDFT (4.3 kcal/mol)\cite{PDFT-MG}. \\

\subsection{Transition States of the Bicyclobutane Isomerization Reaction to gauche-1,3-Butadiene}

\begin{figure}
    \centering
    \includegraphics[scale=0.35]{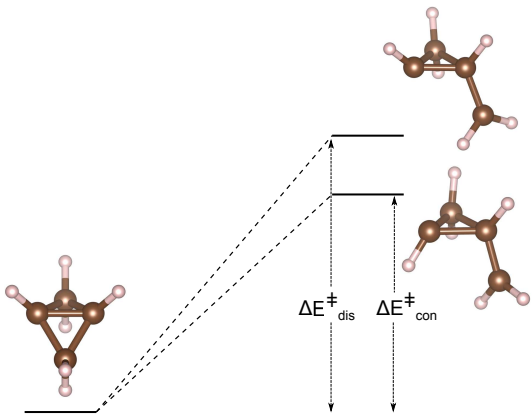}
    \caption{Reaction coordinate of the isomerization reaction of bicyclobutane to 1,3-butadiene with the con- and dis-rotatory transition states shown. The conversion to the 1,3-butadiene is not considered. }
    \label{fig:RXN}
\end{figure}

As a final example, we consider the orbital dependence in the use of CI and post-CI methods to model a simple chemical reaction, calculating the energy barrier, $\Delta H^{\ddagger}$, to the isomerization reaction of bicyclobutane to gauche-1,3-butadiene. This isomerization process may proceed via two different transition states arising from conrotatory (CON) and disrotatory (DIS) pathways. The reaction coordinate diagram displaying these is shown in Figure \ref{fig:RXN}.
Optimized geometries and zero-point and vibrational corrections to the electronic energy were obtained from reference \cite{David_Bicyc}, using the MCSCF/6-31G*\cite{631gs} level of theory, and amounting to 0.0911, 0.0862, and 0.0844~hartrees for bicyclobutane, and the CON and DIS transition states, respectively. Calculations were performed with the previously surveyed methods, CAS orbitals were selected around the HOMO and LUMO, and the results are compared to those obtained from [14,14] CASSCF/ACSE calculations, as well as previously reported ACSE\cite{David_Bicyc}, CC(P;Q)\cite{piecuch_bicyc}, and DMC\cite{DMC_bicyc} data. The barrier for the CON transition pathway has been experimentally determined to be 40.6 $\pm$ 2.5 kcal/mol. The data are shown in Table \ref{tab:DAdata}.\\

\begin{table*}
    \centering
    \begin{tabular}{cc|cc|ccccccccccc}
         & & \multicolumn{2}{c|}{CASSCF} &  \multicolumn{11}{c}{\textcolor{black}{Molecular Orbitals}}   \\
        \textcolor{black}{Pathway} & & [4,4] & [14,14] & HF & MP2 & CCSD & V2-T & LDA & PBE & BLYP & B3LYP & M062X & wB97XD & MN15 \\
        \hline
         \multirow{7}{*}{CON} & $\Delta H^{\ddagger}$ &  &  & 57.59 & 48.60 & 47.53 &  & 41.92 & 42.25 & 36.75 & 43.50 & 53.42 & 50.30 &   50.89 \\
         & $\Delta H^{\ddagger}_{\text{CAS}}$ & 37.68 & 33.70 & 42.86 & 39.49 & 31.16 & 37.85 & 45.92 & 45.23 & 44.51 & 43.07 & 43.25 & 42.91 &  42.80  \\
         & $\Delta H^{\ddagger}_{\text{ACSE}}$ & 40.56 & 40.78 & 46.56 & 46.71 & 43.43 & 40.74 & 39.54 & 42.14 & 42.49 & 43.38 & 43.63 & 44.01 & 42.95 \\
         & $\lambda_{\text{HONO,CAS}}$ & 1.701 & 1.734 & 1.816 & 1.846 & 1.755 & 1.701 & 1.782 & 1.776 & 1.770 & 1.776 & 1.794 & 1.787 & 1.785 \\
         & $\lambda_{\text{LUNO,CAS}}$ & 0.299 & 0.271 & 0.187 & 0.155 & 0.247 & 0.299 & 0.220 & 0.224 & 0.230 & 0.225 & 0.209 & 0.216 & 0.218  \\
         & $\lambda_{\text{HONO,ACSE}}$ & 1.669 & 1.703 & 1.763 & 1.800 & 1.720 & 1.669 & 1.722 & 1.723 & 1.719 & 1.726 & 1.741 & 1.737 & 1.732 \\
         & $\lambda_{\text{LUNO,ACSE}}$ & 0.319 & 0.286 & 0.228 & 0.188 & 0.270 & 0.319 & 0.281 & 0.277 & 0.280 & 0.271 & 0.256 & 0.261 & 0.267 \\
        \hline
         \multirow{7}{*}{DIS} & $\Delta H^{\ddagger}$ & &  & 92.38 & 68.98 & 67.51 &  & 64.45 & 64.46 & 58.18 & 67.69 & 80.78 & 76.36 & 76.95   \\
         & $\Delta H^{\ddagger}_{\text{CAS}}$ & 46.49 & 47.25 & 53.30 & 53.91 & 40.07 & 46.69 & 51.63 & 50.21 & 50.10 & 49.75 & 50.62 & 50.03 & 49.31   \\
         & $\Delta H^{\ddagger}_{\text{ACSE}}$ & 52.11 & 51.79 & 55.93 & 59.59 & 54.32 & 52.33 & 49.35 & 51.76 & 52.43 & 53.53 & 53.15 & 53.53 & 51.72 \\
         & $\lambda_{\text{HONO,CAS}}$ & 1.364 & 1.413 & 1.462 & 1.612 & 1.438 & 1.364 & 1.434 & 1.428 & 1.418 & 1.420 & 1.433 & 1.429 & 1.428 \\
         & $\lambda_{\text{LUNO,CAS}}$ & 0.636 & 0.589 & 0.547 & 0.398 & 0.566 & 0.636 & 0.573 & 0.580 & 0.590 & 0.588 & 0.576 & 0.580 & 0.581  \\
         & $\lambda_{\text{HONO,ACSE}}$ & 1.345 & 1.393 & 1.423 & 1.574 & 1.416 & 1.345 & 1.399 & 1.396 & 1.387 & 1.390 & 1.400 & 1.398 & 1.395 \\
         & $\lambda_{\text{LUNO,ACSE}}$ & 0.641 & 0.594 & 0.567 & 0.419 & 0.574 & 0.641 & 0.602 & 0.605 & 0.612 & 0.608 & 0.598 & 0.601 & 0.603 \\
    \end{tabular}
    \caption{Data for the con- and disrotatory pathways of the isomerization reaction of bicyclobutane. Calculations were carried out with the 6-31G* basis set and CASCI and ACSE calculations utilize a [4,4] active space. Geometries and free energy corrections calculated at the MCSCF/6-31G* level of theory and were obtained from reference\cite{David_Bicyc}. $\Delta H^{\ddagger}$ denotes the transition state barrier \textcolor{black}{in kcal/mol} including zero point and vibrational corrections amount to -3.087 kcal/mol and -4.221 kcal/mol for the CON and DIS pathways, respectively. $\lambda_{\text{HONO}}$ and $\lambda_{\text{LUNO}}$ denote the occupations of the highest and lowest natural orbitals (HONO and LUNO), respectively.}
    \label{tab:DAdata}
\end{table*}

First considering the conrotatory pathway, variation across the barriers predicted by the single reference methods is significant, ranging from $\Delta H^{\ddagger} = 36.75$ kcal/mol with BYLP to $\Delta H^{\ddagger} = 57.59$ kcal/mol with HF. Within the DFT realm, the obtained results are very sensitive to the choice of functional and lack consistency, with BLYP underestimating $\Delta H^{\ddagger}$ while the remaining functionals overestimate it and variations far exceeding chemical accuracy. Nonetheless, the LDA and PBE functionals predict $\Delta H^{\ddagger}$ values that lie within the experimental bounds of error. It is noteworthy that the MN15 functional which is fitted to perform well in multi-reference problems, gives a large overestimation of $\Delta H^{\ddagger}$ with the predicted 50.89 kcal/mol being far outside the realms of chemical accuracy. \\

Using a [14,14] active space CASSCF yields $\Delta H^\ddagger = 33.70$ kcal/mol, with highest occupied natural orbital (HONO) and lowest unoccupied natural orbital (LUNO) occupation numbers of 1.734 and 0.271, respectively, making the CON transition state the less correlated one. The static correlation from the CASSCF calculation overstabilizes the TS compared to bicyclobutane, resulting in underestimation of $\Delta H^{\ddagger,CAS}$. For our CASCI calculations we use a smaller [4,4] active space ($\Delta H^{\ddagger}_{\text{CAS}} = 37.68$), which reduces the magnitude of this underestimation and is sufficient to resolve the biradical character of the TS, yielding HONO and LUNO occupation numbers of 1.701 and 0.299, respectively. For the CASCI calculations, CCSD again provides the orbitals most optimal to resolve the multireference correlation of the methods surveyed, underestimating the CON barrier, and yielding the smallest deviation from the [14,14] CASSCF result and with $\Delta H^{\ddagger} = 31.16$ kcal/mol---a lower barrier than both [4,4] and [14,14] CASSCF. All other orbitals provide a CASCI energy with a positive deviation from the CASSCF $\Delta H^{\ddagger}$, with MP2 NOs yielding the least correlated solution but the smallest error, followed by HF and finally the various DFT functionals, which yield large $\Delta H^{\ddagger}$s but more correlated solutions than MP2 and HF, showing HONO and LUNO occupations numbers comparable to [4,4] CASSCF. \\

Using the ACSE to resolve the full correlation energy, both [4,4] and [14,14] CASSCF resolve the CON barrier to near-exact accuracy providing near-identical results of 40.74 kcal/mol and 40.78 kcal/mol, respectively. MP2 NOs provides both the least correlated solution, as well as the largest deviation from the experimental range of $\Delta H^{\ddagger}$, lying 3.61 kcal/mol above this interval. It is closely followed in both error and correlation by HF. Contrary to the results from the S-T gaps and N$_2$ dissociation, CCSD NOs are now outperformed by the majority of DFT functionals, with only M06-2X, and $\omega$B97XD deviating by more than CCSD's 0.33 kcal/mol from the experimental confidence interval. The LDA, PBE, BLYP and MN15 orbitals all yield $\Delta H^{\ddagger}$ values obtained by the CASCI/ACSE algorithm that lie within the experimental error bound. \\

The disrotatory TS provides for the more correlated and higher energy isomerization pathway, with [14,14] CASSCF/ACSE yielding a barrier of $\Delta H^{\ddagger} = 51.79$ kcal/mol and LUNO and LUNO occupation numbers of 1.393 and 0.594, respectively. There is no experimental reference data for the DIS pathway, but DMC calculations have yielded a barrier of 58.6 kcal/mol\cite{DMC_bicyc}, while CR-CC(2,3) predicts a barrier height of 67.5 kcal/mol\cite{piecuch_bicyc}. Across the various single-reference methods and the CASCI calculations, the trends remain unchanged from the CON pathway, however, with increased errors in the single-reference calculations as the degree of multi-reference correlation in the TS is increased. All CASCI/ACSE calculations with DFT orbitals fall within the $\pm$ 2.5 kcal/mol range of the CASSCF[14,14]/ACSE reference, with MN15 and PBE yielding the closest, and near-identical, results. In this more strongly correlated TS, CCSD NOs yield a larger error of 2.53 kcal/mol. As in the CON TS, DFT orbitals yields more fractional NON than those from HF, MP2, and comparable values to CCSD NOs. \\

\begin{table}
    \centering
    \begin{tabular}{cc|cccc}
         & & \multicolumn{4}{c}{\textcolor{black}{Molecular Orbitals}}\\
         & & M06-L & M06 & M06-2X & M06-HF \\
        \textcolor{black}{Pathway} & \% HF & 0 & 27 & 54 & 100\\
        \hline
        \multirow{7}{*}{CON} & $\Delta H^{\ddagger}$ & 46.72 & 48.46 & 53.42 & 57.46 \\
        & $\Delta H^{\ddagger}_{CAS}$ & 44.87 & 43.16 & 43.25 & 42.66 \\
        & $\Delta H^{\ddagger}_{ACSE}$ & 42.14 & 43.03 & 43.63 & 45.15 \\
        & $\lambda_{\text{HONO,CAS}}$ & 1.785 & 1.787 & 1.794 & 1.806 \\
        & $\lambda_{\text{LUNO,CAS}}$ & 0.215 & 0.214 & 0.209 & 0.198 \\
        & $\lambda_{\text{HONO,ACSE}}$ & 1.732 & 1.735 & 1.741 & 1.755 \\
        & $\lambda_{\text{LUNO,ACSE}}$ & 0.269 & 0.264 & 0.256 & 0.240 \\
        \hline
        \multirow{7}{*}{DIS} & $\Delta H^{\ddagger}$ & 72.43 & 74.69 & 80.78 & 85.41 \\
        & $\Delta H^{\ddagger}_{CAS}$ & 49.29 & 49.60 & 50.62 & 52.26 \\
        & $\Delta H^{\ddagger}_{ACSE}$ & 51.76 & 52.57 & 53.15 & 53.87\\
        & $\lambda_{\text{HONO,CAS}}$ & 1.426 & 1.427 & 1.433 & 1.449 \\
        & $\lambda_{\text{LUNO,CAS}}$ & 0.582 & 0.582 & 0.576 & 0.560 \\
        & $\lambda_{\text{HONO,ACSE}}$ & 1.395 & 1.395 & 1.400 & 1.413 \\
        & $\lambda_{\text{LUNO,ACSE}}$ & 0.606 & 0.604 & 0.598 & 0.580 \\
    \end{tabular}
    \caption{Data for \textcolor{black}{the} con- and disrotatory pathways of the bicylobutane isomerization resolved for the members of the MN06 suite of functionals with their varying degrees of exact HF-exchange. \textcolor{black}{Calculations were carried out with the 6-31G* basis set and CASCI and ACSE calculations utilize a [4,4] active space. Geometries and free energy corrections calculated at the MCSCF/6-31G* level of theory and were obtained from reference\cite{David_Bicyc}. $\Delta H^{\ddagger}$ denotes the transition state barrier in kcal/mol including zero point and vibrational corrections amount to -3.087 kcal/mol and -4.221 kcal/mol for the CON and DIS pathways, respectively. $\lambda_{\text{HONO}}$ and $\lambda_{\text{LUNO}}$ denote the occupations of the highest and lowest natural orbitals (HONO and LUNO), respectively}.}
    \label{tab:M06_exchange}
\end{table}

Lastly, we again look at the M06 suite of functionals to resolve the influence of HF exchange in the DFT functional. While there is no obvious trend in the barrier height predicted by CASCI based on the various orbitals, $\Delta H^{\ddagger}$ predicted by the functional and the CASCI/ACSE calculation, as well as, the NONs follow the expected trend with a lower fraction of HF exchange better accounting for the multi-reference correlation in the studied transition states. Consequently, errors in $\Delta H^{\ddagger}$ and the value of the NON increase across the series from M06-L to M06-HF. The M06-L functional MOs provide a barrier height within the experimental range of error for the CON pathway, yielding identical results to PBE and an only slightly larger error than LDA, while in the DIS pathway M06-L orbitals yield near-exact agreement with [14,14] in both NON and $\Delta H^{\ddagger}$, providing the best orbitals from any method surveyed. \\

\section{Discussion \& Conclusions}

We have employed CASCI calculations in combination with the ACSE to resolve the orbital dependence on the dynamic and multi-reference parts of the total electronic correlation energy. Considering problems dominated by multi-reference correlation, we show that CASCI calculations display significant dependence on the chosen molecular orbital basis, with coupled cluster natural orbitals yielding the most optimal orbitals to account for multi-reference correlation of the single-reference methods surveyed, and HF yielding the least suitable orbitals, while DFT functionals lie between the two methods. Nonetheless, for the accurate prediction of multi-reference dependent properties through the means of CI calculations only, CASSCF orbital optimization is prudent. Use of a post-CI method to account for dynamic correlation, in this case the ACSE, reduces the orbital dependence of the accuracy in the predicted properties. \\

Using the ACSE to resolve post-CI dynamic correlation, we survey orbitals from wave-function based single-reference, as well as, various popular DFT functionals. While HF orbitals yield good results for the N$_2$ dissociation, they tend to fail to capture accurately multi-reference character and deliver lackluster results in the CASCI/ACSE scheme for the prediction of biradical S-T gaps and TS barriers. MP2 is plagued by inconsistencies and convergence issues. Natural orbitals obtained from CCSD calculations, however, allow CASCI/ACSE to resolve both dynamic and strong correlation effects in the three case studies accurately, outperforming CASSCF orbitals in the N$_2$ dissociation, where molecular geometries not dominated by static correlation are considered, most closely mirroring the FCI dissociation curve, and yielding biradical S-T gaps and bicyclobutane isomerization barriers with accuracies close to those achieved with CASSCF orbitals. \\

The various DFT functionals, which are known to yield widely varying results for different systems and properties based on their parametric fitting, produce orbitals that compared to the results predicted by the functionals themselves, such as S-T gaps or dissociation energies, show much greater consistency. Of the tested functionals, the M06 suite and the $\omega$B97XD functionals provide the best suited orbitals for the CASCI/ACSE calculations, yielding only marginally worse performance than CASSCF and CCSD orbitals. Furthermore, resolving the S-T gaps and bicyclobutane isomerization barriers obtained with the M06 suite functionals shows the most optimal orbitals to account for both multi-reference and dynamic correlation are obtained with the lowest HF-exchange fraction, i.e. the M06-L functional, which yields the best orbitals for the treatment of multi-reference problems of any tested functional. As DFT presents the most ubiquitous electronic structure method across many disciplines in chemistry, physics and materials science, implemented in any commonly used software package, offering inexpensive computational scaling compared to CASSCF or CC methods, it provides a good compromise between computational costs, ease-of-use, and accuracy. Especially, considering the fact that DFT molecular orbitals are already available in most cases through prior geometry optimizations or frequency calculations, they provide a viable option for further ab-initio calculations aimed at resolving electron correlation in many applications, significantly reducing further computational expense while retaining viable accuracy. \\

This work provides valuable insight into the orbital dependence in the ability of CASCI and post-CI methods to resolve multi-reference and dynamic correlation. We demonstrate that contrary to popular implementations that rely on CASSCF orbital optimizations for the resolution of the total correlation energy, CASSCF may not always provide the optimal molecular orbital basis set to account for the combination of static and dynamic contributions to the electronic energy. Furthermore, improved computational scaling may be obtained through the use of widely available single-reference methods for the optimization of the molecular orbitals. Additionally, if a post-CI method to resolve all-electron correlation were to be implemented in a SCF fashion, undergoing further orbital optimization after the initial CAS seed calculation, performance of an initial CASSCF calculation may be of limited value as compared to a seed with orbitals obtained from the surveyed single-reference methods. Throughout the studied systems we show that the CASCI/ACSE method is a valuable tool in the accurate resolution of the properties of a multi-reference system, and may be used in combination with any single-reference calculation, in particular with DFT, not requiring further CASSCF calculations. \\

\begin{acknowledgments}
D.A.M. gratefully acknowledges the U.S. National Science Foundation Grant CHE-1565638.
\end{acknowledgments}


%

\end{document}